\documentclass[12pt]{article}

\usepackage[english]{babel}
\usepackage[T1]{fontenc}
\usepackage[utf8]{inputenc}
\usepackage{amssymb,amsmath,array,hhline,bm,curves}
\usepackage{latexsym}
\usepackage{graphicx}
\usepackage[final]{pdfpages}
\usepackage{verbatim}
\usepackage{enumerate,bbm}

\title{Krein signatures of transfer operators \\ for half-space topological insulators}

\author{Hermann Schulz-Baldes$^{1}$, Carlos Villegas-Blas$^2$
\\
\\
{\small $^1$ Department Mathematik, Friedrich-Alexander-Universit\"at Erlangen-N\"urnberg, Germany}
\\
{\small $^2$ Instituto de Matematicas, Cuernavaca, UNAM, Mexico}
}

\date{ }

\newtheorem{theo}{Theorem}

\newtheorem{proposi}{Proposition}

\newtheorem{coro}{Corollary}

\newcommand{\CM}{{\mathbb C}}
\newcommand{\NM}{{\mathbb N}}
\newcommand{\RM}{{\mathbb R}}
\newcommand{\BM}{{\mathbb B}}
\newcommand{\SM}{{\mathbb S}}
\newcommand{\TM}{{\mathbb T}}

\newcommand{\ZM}{{\mathbb Z}}

\newcommand{\GM}{{\mathbb G}}
\newcommand{\GUM}{{\mathbb G}{\mathbb U}}

\newcommand{\UM}{{\mathbb U}}

\newcommand{\Oo}{{\cal O}}

\newcommand{\Cc}{{\cal C}}

\newcommand{\Kk}{{\cal K}}
\newcommand{\Hh}{{\cal H}}

\newcommand{\JF}{J}
\newcommand{\JR}{S}

\newcommand{\IS}{S_{\mbox{\rm\tiny tr}}}
\newcommand{\KPH}{S_{\mbox{\rm\tiny ph}}}

\newcommand{\etaR}{\eta}
\newcommand{\etaFR}{\tau}

\newcommand{\etaPH}{\eta_{\mbox{\rm\tiny ph}}}

\newcommand{\one}{{\bf 1}}

\newcommand{\Uu}{\mathcal{U}}

\newcommand{\Ch}{\mbox{\rm Ch}}

\DeclareMathOperator{\e}{e}

\newcommand{\Sig}{{\mbox{\rm Sig}}}
\newcommand{\Sec}{{\mbox{\rm Sec}}}

\newcommand{\ph}{{\mbox{\rm\tiny ph}}}

\begin{document}

\maketitle

%%%%%%%%%%%%%%%%%%%%%%%%%%%%%%%%%%%%%%%%%%%%%%%%%%%%
\begin{abstract}
We propose a complementary point of view on the topological invariants of two-dimensional tight-binding models restricted to half-spaces. The transfer operators for such systems are $J$-unitary on a infinite dimensional Krein space $(\Kk,J)$ and, for energies in the bulk gap, only have discrete spectrum on the unit circle. These eigenvalues have Krein inertia which can be used to define topological invariants determining the nature of the surface modes and allowing to distinguish different topological phases. This is illustrated by numerical results.
\end{abstract}
%%%%%%%%%%%%%%%%%%

\vspace{.5cm}

%%%%%%%%%%%%%%%%%%%%%%%%%%%%%%%%%%%%%%%%%%%%%%%%%%%%%%%%%%%%%%%%%%%%%%%%%
\section{Introduction}
\label{sec-intro}

Edge states of quantum systems, also called Tamm states \cite{Tam} or Shockley states \cite{Sho}, have been known for a long time. More recently they started playing a prominent role in the theory of quantum Hall systems \cite{Hal,Hat} and topological insulators \cite{KM,SRFL}. One way to characterize these latter solid state materials is by the existence of edge states that are not susceptible to Anderson localization. The topological protection of the extended edge modes is linked to the non-trivial topological invariants. These invariants have been extensively studied in the mathematical physics literature \cite{ASV,GP,PS}. Here we present an alternative perspective on topological edge states of half-space tight-binding models by analyzing the spectral theory of the transfer operators describing the formal solutions of the associated Schr\"odinger equation on a half-space. In contradistinction to most prior works \cite{Hat,ASV,GP,AEG}, these transfer operators act parallel to the boundary, and {\em not} perpendicular to it. They act on infinite dimensional Hilbert spaces and this opens the way for a much richer spectral theory.

\vspace{.2cm}

Let us briefly describe which type of spectral properties we are investigating below and what their physical relevance is. The parallel transfer operators described above conserve a quadratic form $J$ on an infinite dimensional Hilbert space. Therefore they are so-called $J$-unitary linear operators if the Hilbert space equipped with $J$ is viewed as a Krein space. For energies in the bulk gap of these systems (namely in a gap of the Hamiltonian without a boundary), the parallel transfer operators moreover only have eigenvalues of finite multiplicity on the unit circle and no essential spectrum touching the unit cirle -- except in the non-generic situation of a flat band of surface states. This reflects that almost all formal solutions constucted with these parellel transfer operators are exponentially growing, which in the language of quasi-one-dimensional systems corresponds to closed or evanescent channels. On the other hand, the finite number of eigenvalues on the unit circle lead to extended edge modes along the boundary of the system. This connection is proved in Section~\ref{sec-transferhalf} by using another different set of transfer matrices, namely those in the direction perpendicular to the boundary. The sign of the group velocity of these edge states is then shown to be equal to the Krein signature of the corresponding eigenvalue of the parallel transfer operator (as defined in  \cite{SB,SV} and reviewed in Section~\ref{app}). This provides a clear physical interpretation of the Krein signatures in the present context. Now it turns out that the sum of all these signatures is a topological invariant that is remarkably stable under homotopic deformations of the system which do not close the bulk gap (see Section~\ref{sec-transferhalf}). Hence it can be used to distinguish different phases of the system. The global invariant is even sufficiently stable to also allow for certain types of random perturbations (Section~\ref{sec-rand}). In the case of a quantum Hall system, the invariant can be shown to be equal to the Hall condunctance.

\vspace{.2cm}

Up to now only systems without further symmetries were described. Following the symmetry analysis of topological insulators \cite{SRFL} let us next consider Hamiltonians with either time-reversal symmetry or particle-hole symmetry. Then the parallel transfer operators also has a further symmetry property on the Krein space. These new symmetries involve a complex conjugation and hence specify a real structure on the Krein structure. For classes of transfer operators respecting these symmetries, it is possible to define secondary $\ZM_2$-invariants. The general mathematical analysis of these new invariants is described in detail in the companian paper~\cite{SV} which is summarized in Section~\ref{app}. While this summary is complete (but without proofs), it may nevertheless be useful for the interested reader to also consult~\cite{SV}. These mathematical results are then applied in Sections~\ref{sec-TRIop} and \ref{sec-PHop} to parallel transfer operators with symmetries as obtained from Hamiltonians with extra symmetries. For the Kane-Mele model \cite{KM} with odd time-reversal symmetry, a non-trivial value of the $\ZM_2$-invariant guarantees the existence of helical edge states, while in the case of Bogoliubov-de Gennes Hamiltonians with even particle-hole symmetry it implies the presence of Majorana edge modes. These are two of the examples dealt with in detail in Section~\ref{sec-examples}, another one is the Harper model with a rational flux.

\vspace{.2cm}

Of course, it is also possible to consider systems which have both time-reversal and particle-hole symmetry. However, such systems inherit a chiral symmetry and for such chiral systems the Krein theory of the parallel transfer operators is not interesting. Hence these systems are only briefly considered in this work.

 %%%%%%%%%%%%%%%%%%%%%%%%%%%%%%%%%%%%%%%%%%%%%%%%
\section{Krein signatures for essentially $\SM^1$-gapped $\JF$-unitaries}
\label{app}

This section reviews those results from \cite{SB,SV} which are relevant for the study of transfer operators. All proofs can be found in these references. A Krein space $(\Kk,\JF)$ is a Hilbert space $\Kk$ equipped with a self-adjoint unitary $\JF$, also called the fundamental symmetry. For sake of concreteness it will be chosen to be of the form $\JF=\binom{\one\;\;\;0}{0\;-\one}$ with blocks of equal dimension. The set of $\JF$-unitary operators on the Krein space is a subset of the bounded linear operators $\BM(\Kk)$ on $\Kk$:
\begin{equation}
\label{eq-Junit}
\UM(\Kk,\JF)
\;=\;
\{T\in\BM(\Kk)\,:\,T^*\JF T=\JF\}
\;.
\end{equation}
A $\JF$-unitary operator is called essentially $\SM^1$-gapped if it only contains discrete spectrum on $\SM^1$ (isolated eigenvalues of finite multiplicity), and $\GM\UM(\Kk,\JF)$ denotes the set of these operators. For each eigenvalue $\lambda\in\SM^1$ of $T\in\GM\UM(\Kk,\JF)$, the restriction of $J$ to its generalized eigenspace is known to be non-degenerate. The number of positive and negative eigenvalues of this restriction are the Krein inertia and are denoted by $\nu_+(\lambda)$ and $\nu_-(\lambda)$ respectively. The signuature of $\lambda\in\SM^1$ as eigenvalue of $T$ is then defined to be $\Sig(\lambda,T)=\nu_+(\lambda)-\nu_-(\lambda)$, and the global signature of $T\in\GM\UM(\Kk,\JF)$ is 
\begin{equation}
\label{eq-sigdef}
\Sig(T)
\;=\;
\sum_{\lambda\in\SM^1} \;\Sig(\lambda,T)
\;.
\end{equation}
The main fact about the signature is its continuity w.r.t. the operator norm topology. Next we want to implement further symmetries on the $J$-unitaries. This requires the use of a real structure on $\Kk$, namely an anti-linear involution $\Cc:\Kk\to\Kk$. As for complex number, we will use the notation $\overline{v}=\Cc v$ for $v\in\Kk$ and  $\overline{A}=\Cc A\Cc$ for $A\in\BM(\Kk)$. An operator $A\in\BM(\Kk)$ is then called real if $A=\overline{A}$. Furthermore, $A^t=(\overline{A})^*$ denotes the transposed operator.  A Real Krein space of kind $(\etaR,\etaFR)\in\{-1,1\}^2$ is a complex Krein space $(\Kk,\JF)$ with $\JF=\overline{\JF}$ together with a second real symmetry operator $\JR=\overline{\JR}$ satisfying 
$$
\JR^2\;=\;\etaR\,\one\;,
\qquad
\JF\,\JR\,=\,\etaFR\,\JR\,\JF
\;.
$$
After an adequate orthogonal basis transformation which conserves $\JF$, the real symmetry operator $\JR$ for the kinds $(\etaR,1)$ and $(\eta,-1)$  respectively is, in the grading of $\JF$, of the form
\begin{equation}
\label{eq-Schoice}
\JR
\;=\;
\begin{pmatrix}
\JR_+ & \;\;0 \\ 0 & \JR_-
\end{pmatrix}
\;\;\;
\mbox{\rm with }\;(\JR_\pm)^2=\etaR\,\one
\;,
\qquad
\JR
\;=\;
\begin{pmatrix}
0 & \eta\,\one \\ \one & 0
\end{pmatrix}
\;.
\end{equation}
On a Real Krein space the set of $\JF$-unitaries with Real symmetry $\JR$ is defined by
\begin{equation}
\label{eq-RSym}
\UM(\Kk,\JF,\JR)
\;=\;
\left\{
T\in\UM(\Kk,\JF)\,:\,\JR^*\,\overline{T}\,\JR=T\right\}
\;.
\end{equation}
Furthermore, let us set $\GM\UM(\Kk,\JF,\JR) =\UM(\Kk,\JF,\JR) \cap \GM\UM(\Kk,\JF)$. For $T\in \GM\UM(\Kk,\JF,\JR)$, one now has the following homotopy invariants:

\begin{enumerate}[\rm (i)]

\item If $(\etaR,\etaFR)=(1,1)$, $\Sig(T)\in\ZM$ and a secondary $\ZM_2$-invariant defined by
\begin{equation}
\label{eq-SecDef}
\Sec(T)
\;=\;
\Sig(1,T)\,\mbox{\rm mod}\,2
\;\in\;\ZM_2\;.
\end{equation}

\item If $(\etaR,\etaFR)=(-1,-1)$, $\Sig(T)=0$ and a secondary $\ZM_2$-invariant defined by
\begin{equation}
\label{eq-Sig2Def}
\Sig_2(T)
\,=\,
\Big(
\tfrac{1}{2}\sum_{\lambda\in\SM^1}
\bigl(\nu_+(\lambda)+\nu_-(\lambda)\bigl)\Big)
\;\mbox{\rm mod }2
\;.
\end{equation}

\item If $(\etaR,\etaFR)=(-1,1)$, $\Sig(T)\in 2\,\ZM$.

\item If $(\etaR,\etaFR)=(1,-1)$, $\Sig(T)=0$.

\end{enumerate}

We now have to make the above results applicable to the transfer operators for which the natural fundamental symmetry is $I=\binom{0\;-\one}{\one\;\;\;0}$. It satisfies $I^2=-\one$ and hence $\Kk$ equipped with $I$ is not a Krein space in the above sense. Nevertheless, one can define $\UM(\Kk,I)$ to be the set of operators $T\in\BM(\Kk)$ satisfying $T^* IT=I$. Now the Cayley transform sends $J$ to $I$:
\begin{equation}
\label{eq-Cayley}
C^*\,J\,C\;=\;\imath\,I\,
\;,
\qquad
C
\;=\;
\frac{1}{\sqrt{2}}\,
\begin{pmatrix}
\one & -\imath\,\one \\ \one & \imath\,\one
\end{pmatrix}
\;.
\end{equation}
Therefore $C^*\,\UM(\Kk,J)\,C=\UM(\Kk,I)$. One concludes that it is of no importance whether the fundamental symmetry squares to $\one$ or $-\one$ and one can furthermore check that the values of the Krein signatures are not changed either. This is not true anymore when a Real symmetry is involved and therefore we extend the notion of a kind of a Real Krein space as follows. A Real Krein space $(\Kk,J,S)$ of kind $(\kappa,\eta,\tau)\in\{-1,1\}^{\times 3}$ consists of two real unitaries $J$ and $S$ satisfying $J^2=\kappa\,\one$, $S^2=\eta\,\one$ and $JS=\tau\,SJ$. Again $\UM(\Kk,J,S)$ is defined as in \eqref{eq-RSym} and the invariants can be defined as in (i)-(iv). Hence there is the sign $\kappa$ as a supplementary information. Actually, this does not lead to anything essentially new because one has
\begin{equation}
\label{eq-switch}
C^*\,\UM(\Kk,J,S)\,C
\;=\;
\UM(\Kk,e^{\imath\theta}\,C^*J C,e^{\imath\theta'}\, C^tS C)
\;=
\;\UM(\Kk,J',S')
\;,
\end{equation}
where $e^{\imath\theta}$ and $e^{\imath\theta'}$ are phases chosen such that $J'=e^{\imath\theta}\,C^*J C$ and $S'=e^{\imath\theta'}\, C^tS' C$ are real. In particular, for $J=\binom{\one \;\;0}{0\;-\one}$ one has $J'=\binom{0\;-\one}{\one\;\;\;0}=I$ by \eqref{eq-Cayley}. Now if $(\Kk,J,S)$ is of kind $(\kappa,\eta,\tau)$, the kind of $(\Kk,J',S')$ can be checked to be given by
$$
(\kappa',\eta',\tau')\;=\;(-\kappa,\eta,-\tau)
\;.
$$
Hence the $8$ possible choices of signs only lead to $4$ different subgroups of $\UM(\Kk,J)$. This information is also contained in Table~1 which also lists the classical groups associated to each of these $4$ choices. These classical groups can be read off the identities 
\begin{eqnarray*}
& & \mbox{\rm O}(N,N)\;=\;\UM(\CM^{2N},J,\one)
\;=\;
C\,\UM(\CM^{2N},I,J)\,C^*
\;,
\\
& & 
\mbox{\rm SP}(2N,\RM)
\;=\;
\UM(\CM^{2N},I,\one)
\;=\;
C^*\,\UM(\CM^{2N},J,K)\,C\;,
\\
& & 
\mbox{\rm SO}^*(2N)
\;=\;
\UM(\CM^{2N},I,I)
\;=\;
C^*\,\UM(\CM^{2N},J,I)\,C
\;,
\\
& &
\mbox{\rm SP}(2N,2N)
\;=\;
\UM(\CM^{4N},J\otimes\one,J\otimes I)
\;=\;
C\otimes \one\,\UM(\CM^{4N},I\otimes\one,K\otimes I)\,C^*\otimes\one
\;,
\end{eqnarray*}
where $K=\binom{0\;\one}{\one\;0}$. Table~1 also contains the corresponding invariants as defined in (i)-(iv).

%\begin{table*}[t]
\begin{table*}
%\begin{table}
\label{table1}
\begin{center}
\begin{tabular}{|c|c||c||c|c||c|c|}
\hline
$(\kappa,\eta,\tau) $ & $(\kappa',\eta',\tau')$ & Class. Group & $\pi_0\supset$ & Invariant  & System & Effect
\\
\hline\hline
$(1,1,1)$ & $(-1,1,-1)$  & $\mbox{\rm O}(N,N)$ & $\ZM\times\ZM_{2}$ & $\Sig\times\Sec$ & even PHS & TQHE
\\
\hline
$(1,1,-1)$ & $(-1,1,1)$ &  $\mbox{\rm SP}(2N,\RM)$ & $1$ & & even TRS & 
\\
\hline
$(1,-1,-1)$ & $(-1,-1,1)$   & $\mbox{\rm SO}^*(2N)$ & $\ZM_2$ & $\Sig_2$ & odd TRS & QSHE
\\
\hline
$(1, -1,1)$ & $(-1,-1,-1)$  & $\mbox{\rm SP}(2N,2N)$ & $2\,\ZM$ & $\Sig$ & odd PHS & SQHE
\\
\hline
\end{tabular}
\caption{\sl 
Table of changes of the kind under the Cayley transform as in \eqref{eq-switch}. Next follow the classical groups that $\UM(\Kk,\JF,\JR)$ is isomorphic to when $\Kk=\CM^{2N}$ or $\CM^{4N}$.  Furthermore are listed the minimal number of connected components $\pi_0=\pi_0(\GUM(\Kk,\JF,\JR))$ of the associated essentially $\SM^1$-gapped operators as well as the global invariants labeling these components. Finally appear the symmetries for which the transfer operators lie in the corresponding $\UM(\Kk,\JF,\JR)$ and the physical effect linked to the invariants (see Sections~\ref{sec-TRIop} and \ref{sec-PHop}). This agrees with \cite{LSS}.
}
\end{center}
\end{table*}

%%%%%%%%%%%%%%%%%%%%%%%%%%%%%%%%%%%%%%%%%%%%%%%%
\section{Periodic Hamiltonians}
\label{sec-periodic}

Let us begin by presenting the tight-binding Hamiltonians studied in this work. They act on the Hilbert space $\Hh=\ell^2(\ZM^2)\otimes\CM^L$. Here the fibers $\CM^L$ allow to describe spin degree of freedoms, particle-hole spaces as for BdG Hamiltonians and sublattice degrees of freedom as needed for the honeycomb lattice. In this fiber, the Real symmetry will be implemented further below. On the Hilbert space $\Hh$ act the two shift operators $S_1$ and $S_2$ defined by $(S_1\phi)_{n_1,n_2}=\phi_{n_1-1,n_2}$ and $(S_2\phi)_{n_1,n_2}=\phi_{n_1,n_2-1}$, as well as the components of the position operator $X_1$ and $X_2$ (densely) defined by $(X_1\phi)_{n_1,n_2}=n_1\phi_{n_1,n_2}$ and $(X_2\phi)_{n_1,n_2}=n_2\phi_{n_1,n_2}$. Then the magnetic shifts at flux $\varphi\in[0,2\pi)$ are, in Landau gauge, given by $U_1=e^{-\imath\varphi X_2}S_1$ and $U_2=S_2$. The shifts commute, but the magnetic translations satisfy the relation $U_1U_2=e^{\imath \varphi}U_2U_1$. Let us also introduce $U_3=U_1^*U_2$ and $U_4=U_1U_2$ describing the next to next nearest neighbor hopping terms. We consider translation invariant (self-adjoint) Hamiltonians of the form
\begin{equation}
\label{eq-genH}
H\;=\;\sum_{i=1,2,3,4}(W_i^*U_i+W_iU_i^*)\;+\;V
\;,
\end{equation}
where the $W_i$ and $V=V^*$ are square matrices of size $L$. As explained in detail in \cite{ASV}, Laplacians on the triangular and honeycomb lattice can be written in this form.  Actually, \eqref{eq-genH} slightly extends the class of matrix-valued periodic operators used  in \cite{ASV}. The Hamiltonian $H$  can be viewed as a periodic Jacobi operator with operator symbols:
\begin{equation}
\label{eq-HamJac1}
H
\; = \;
A \,S_1^*+B+A^*\,S_1
\;,
\end{equation}
where the coefficient operators acting on $\ell^2(\ZM)\otimes \CM^L$ are given by
\begin{equation}
\label{eq-ABdef}
A\;=\;
W_1\,e^{\imath\varphi X_2}\,+\,W_3^*\,e^{\imath\varphi X_2}\, S_2\,+\,W_4\,S_2^*\,e^{\imath\varphi X_2}\;,\qquad
B\;=\;W_2^*S_2+W_2S_2^*+V
\;.
\end{equation}
The Landau gauge was chosen such that the Hamiltonian can be written in the form \eqref{eq-HamJac1} of a periodic Jacobi matrix with operator valued entries $A$ and $B$. This immediately allows to apply transfer operator techniques developed in Section~\ref{sec-transfer} below. Furthermore, one can readily partially diagonalize $H$ by the partial Fourier transform $\Uu_1:\ell^2(\ZM)\to L^2(\TM^1,\frac{dk_1}{2\pi})$ defined by
\begin{equation}
\label{eq-Fourier1}
(\Uu_1{\psi})(k_1)\;=\;
\sum_{n_1\in\ZM}{\psi}_{n_1}\,e^{\imath n_1k_1}\;,
\qquad
k_1\in\TM^1=(-\pi,\pi]\;,
\end{equation}
and naturally extended if there are fibers. For example, one obtains a unitary $\Uu_1: \ell^2(\ZM^2)\otimes\CM^L\to L^2(\TM^1,\frac{dk_1}{2\pi})\otimes\ell^2(\ZM)\otimes\CM^{L}$. The inverse is given by
$$
(\Uu_1^*\phi)_{n_1}\;=\;
\int_{-\pi}^\pi\frac{d k_1}{2\pi}\;\,\phi(k_1)\,\e^{-\imath k_1n_1}\;.
$$
With this unitary, one has
\begin{equation}
\label{eq-Hfibered}
\Uu_1\,H\,\Uu_1^*\;=\;\int^\oplus_{\TM^1} dk_1\;H_2(k_1)
\;,
\end{equation}
with a self-adjoint fiber operator $H_2(k_1)$ acting on $\ell^2(\ZM)\otimes \CM^L$ and given by
$$
H_2(k_1)
\;=\;
A\,e^{-\imath k_1}+B+A^*\,e^{\imath k_1}
\;.
$$
Each $H_2(k_1)$ is again a Jacobi matrix, now with matrix entries in the $L\times L$ matrices.  If the magnetic flux  vanishes, then it is $1$-periodic and can be diagonalized by the Fourier transform $\Uu_2$ defined exactly as $\Uu_1$. Setting $\Uu=\Uu_1\Uu_2$, one has, still for $\varphi=0$, that $\Uu H\Uu^*=\int^\oplus_{\TM^2}d{\bf k}\; H({\bf k})$ where with ${\bf k}=(k_1,k_2)$
$$
H({\bf k})
\;=\;
\sum_{i=1,2,3,4}(\e^{\imath k_i}W_i^*+
\e^{-\imath k_i}W_i)
+V
\;,
\qquad
k_3=k_2-k_1\;,\;\;k_4=k_2+k_1
\;.
$$
Each $H({\bf k})$ is an $L\times L$ matrix and its $L$ eigenvalues form the so-called Bloch bands of $H$. 

\vspace{.2cm}

If $\frac{\varphi}{2\pi}=\frac{q}{p}$ is rational, then $H$ and the Jacobi matrix $H_2(k_1)$ are $p$-periodic in the $2$-direction and can be diagonalized by the Bloch-Floquet transformation which will also be denoted by $\Uu_2$ as it generalizes the above. It sends $\ell^2(\ZM)$ to $L^2(\TM^1_p,\frac{dk_2}{|\TM^1_p|})\otimes \ell^2(\ZM_p)$ where $\TM^1_p=(-\frac{\pi}{p},\frac{\pi}{p}]$ and $\ZM_p=\{0,1,\ldots,p-1\}$, and it is given by
$$
(\Uu_2\phi)_{l}(k_2)
\;=\;
\sum_{m\in\ZM}\phi_{l+pm}\;e^{\imath(l+pm) k_2}
\;,
\qquad
l\in\ZM_p\;,
\;\;
k_2\in\TM^1_p
\;.
$$
Again $H$ is diagonalized by $\Uu=\Uu_1\Uu_2$, but now $H({\bf k})$ is a $Lp\times Lp$ matrix. Furthermore, $\Uu_2H\Uu_2^*$ is fibered similarly as in \eqref{eq-Hfibered}, but with fibers $H_1(k_2)$ which are Jacobi matrices with matrix entries of size $Lp\times Lp$. The spectrum of each $H_2(k_1)$ and $H_1(k_2)$ is typically absolutely continuous, but it may happen that flat bands given by infinitely degenerate eigenvalues appear. For irrational $\frac{\varphi}{2\pi}$, one is confronted with a quasiperiodic operator (an example being the Harper Hamiltonian with irrational flux, then $k_1$ is the phase). This case is not considered here. Let us note that also $\Uu_2A\Uu_2^*$ and $\Uu_2B\Uu_2^*$ are fibered operators with fibers given by $Lp\times Lp$ matrices $A(k_2)$ and $B(k_2)$ respectively. For the sake of completeness, let us write them out explicitly. For $m,l\in\ZM_p$ and with Kronecker delta's modulo $p$,
\begin{eqnarray}
\langle m|A(k_2)|l\rangle
& = &
W_1\,e^{\imath\varphi m}\,\delta_{m,l}
\;+\;
W_3\,e^{\imath(\varphi m-k_2)}\,\delta_{m,l-1}
\;+\;
W_4^*\,e^{\imath(\varphi m+\varphi+k_2)}\,\delta_{m,l+1}
\;,
\nonumber
\\
\label{eq-ABk2}
\\
\langle m|B(k_2)|l\rangle
& = & 
W_2^*\,e^{-\imath k_2}\,\delta_{m,l-1}
\;+\;
W_2\,e^{\imath k_2}\,\delta_{m,l+1}
\;+\;
V
\;.
\nonumber
\end{eqnarray}

%%%%%%%%%%%%%%%%%%%%%%%%%%%%%%%%%%%%%%%%%%%%%%%%
\section{Half-space restrictions of periodic Hamiltonians}
\label{sec-restrictperiodic}

In this section, the physical space is a half-space $\ZM\times\NM$ where the $1$-direction is $\ZM$ and the $2$-direction is $\NM$ (with $0$ incluced). The half-plane operators $\widehat{H}$ are obtained by restricting $H$ given by \eqref{eq-genH} to the half-space $\ZM\times\NM$ so that the Hilbert space is $\widehat{\Hh}=\ell^2(\ZM\times\NM)\otimes\CM^L$. For simplicity, the restriction is simply obtained by imposing Dirichlet boundary conditions. This means that the shift $S_2$ is replaced by a partial isometry $\widehat{S}_2$ defined by $(\widehat{S}_2)^*\widehat{S}_2=\one-|0\rangle\langle 0|$ and $\widehat{S}_2(\widehat{S}_2)^*=\one$, while $\widehat{S}_1=S_1$. Then one defines $\widehat{U}_j$, $j=1,2,3,4$, as above and using this restricted magnetic translations the Hamiltonian $\widehat{H}$ is given by the same formula \eqref{eq-genH}. Furthermore, also operators $\widehat{A}$ and $\widehat{B}$ are defined by restriction, notably these are operators on $\ell^2(\NM)\otimes\CM^L$. 

\vspace{.2cm}

The operator $\widehat{H}$ is still translation invariant in the $1$-direction and can hence be partially diagonalized by $\Uu_1$:
$$
\Uu_1 \widehat{H}\,\Uu_1^*
\;=\;
\int^\oplus_{\TM^1}dk_1\; \widehat{H}_2(k_1)
\;,
$$ 
where now $\widehat{H}_2(k_1)$ is an operator on $\ell^2(\NM)\otimes\CM^{L}$ given by
\begin{equation}
\label{eq-H2k1}
\widehat{H}_2(k_1)
\;=\;
\widehat{A}\,e^{-\imath k_1}+\widehat{B}+\widehat{A}^*\,e^{\imath k_1}
\;.
\end{equation}
Using the restriction of the formulas \eqref{eq-ABdef}, one readily writes out an explicit formula for  $\widehat{H}_2(k_1)$ and this will be done in Section~\ref{sec-calcsig}. It is a one-sided Jacobi matrix with $L\times L$ matrix entries. If $\frac{\varphi}{2\pi}$ is rational, then its coefficients are periodic and, if $\varphi=0$, then the coefficients are constant. The one-sided Jacobi matrix $\widehat{H}_2(k_1)$ can have bound states $E^{\mbox{\rm\tiny e}}_n(k_1)$ depending analytically on $k_1$. These so-called edge bands constitute the edge spectrum $\sigma^{\mbox{\rm\tiny e}}(\widehat{H})$ of $\widehat{H}$ consisting by definition \cite{DS,ASV} of those $E\in\RM $ for which there is a $e^{\imath k_1}\in\TM^1$ such that $E$ is an eigenvalue of $\widehat{H}_2(k_1)$. Albeit untypical, it may happen that the eigenvalues $E^{\mbox{\rm\tiny e}}_n(k_1)$ are constant in $k_1$. This then provides a so-called flat band of edge states.

%%%%%%%%%%%%%%%%%%%%%%%%%%%%%%%%%%%%%%%%%%%%%%%%
\section{Transfer operators of the planar models}
\label{sec-transfer}

The transfer operators allow to construct formal solutions of the Schr\"odinger equation $H\psi=E\psi$ associated to the Hamiltonian \eqref{eq-HamJac1}. To avoid technicalities, let us first assume  the following hypothesis which will be discussed and weakened further below.

\vspace{.2cm}

\noindent {\bf Strong Invertibility Hypothesis:} {\sl $A$ given in} \eqref{eq-ABdef} {\sl is invertible with bounded inverse.} 

\vspace{.2cm}

\noindent Then one can introduce the transfer operator in the $1$-direction and at energy $E\in\RM$ by
\begin{equation}
\label{eq-transferopdef}
T^E_1
\;=\;
\begin{pmatrix}
(E\,{\bf 1}-B)A^{-1}
& -\,A^*
\\
A^{-1} & 0
\end{pmatrix}\;.
\end{equation}
It acts on the Hilbert space $\Kk=\ell^2(\ZM)\otimes\CM^L\otimes\CM^2$. It can then readily be checked that the transfer operator satisfy
\begin{equation}
\label{eq-transferopfundsym}
(T^E_1)^*\,I\,T^E_1
\;=\;
I\;,
\qquad
I\;=\;
\begin{pmatrix}
0 & -\,\one
\\
\one & 0
\end{pmatrix}\;.
\end{equation}
Therefore, in the terminology of Section~\ref{app}, $T^E_1$ is an $I$-unitary on the Krein space $(\Kk,I)$. This property is often referred to as a symplectic symmetry, but really does not involve a real structure yet. If $\psi=(\psi_{n})_{n\in\ZM}$ with fiber vectors $\psi_n\in\ell^2(\ZM)\otimes\CM^L$, then the solution of the Schr\"odinger equation (not necessarily square summable in $n$) satisfies 
\begin{equation}
\label{eq-transferrel}
\begin{pmatrix}
A\psi_{n+1} \\ \psi_{n}
\end{pmatrix}
\;=\;
T_1^E\,
\begin{pmatrix}
A\psi_{n} \\ \psi_{n-1}
\end{pmatrix}\;.
\end{equation}

\vspace{.2cm}

Also the transfer operators are diagonalized by partial Fourier transforms:
\begin{equation}
\label{eq-transferFourier}
\Uu_2\;T_1^E\;\Uu_2^*
\;=\;
\int^\oplus_{\TM^1_p} dk_2\;T_1^E(k_2)
\;,
\end{equation}
with $2Lp\times 2Lp$ matrices given by
$$
T_1^E(k_2)
\;=\;
\begin{pmatrix}
(E\,{\bf 1}-B(k_2))A(k_2)^{-1}
& -\,A(k_2)^*
\\
A(k_2)^{-1} & 0
\end{pmatrix}\;,
$$
with $A(k_2)$ and $B(k_2)$ given by \eqref{eq-ABk2}. The matrices $T_1^E(k_2)$ are $I$-unitary and analytic in $k_2$. Hence their spectrum is given by $Lp$ curves which are locally analytic and have Puiseux expansions at level crossings. These curves constitute the spectrum of the multiplication operator $\Uu_2T_1^E\Uu_2^*$ which is hence absolutely continuous. In most examples known to us \cite{ASV} as well as those dealt with in Section~\ref{sec-examples} the above strong hypothesis holds. Furthermore, adding a generic small perturbation to the Hamiltonian (more precisely, some further small hopping between the neighboring sites) will make the Strong Invertibility Hypothesis valid. Hence, on physical grounds there is no reason not suppose it to hold as effects resulting from the singularity of $A$ would not be structurally stable. However, in some particular models such as the Laplacian on the honeycomb lattice (without staggered potential and spin orbit coupling as in \cite{ASV}) only the following holds true.

\vspace{.2cm}

\noindent {\bf Weak Invertibility Hypothesis:} {\sl $A(k_2)$ is invertible for almost all $k_2$.} 

\vspace{.2cm}

\noindent Then $A^{-1}$ and thus $\Uu_2T_1^E\Uu_2^*$ are clearly densely defined and the above spectral analysis by Fourier transform still applies. As this spectrum does not cover all $\CM$, it follows that $\Uu_2T_1^E\Uu_2^*$ is closed by Lemma 1.1.2 of \cite{Dav}. In conclusion, the weak hypothesis guarantees that $T_1^E$ is $I$-isometric in the sense of \cite{SV}. Therefore all results of this chapter readily carry over to systems which satisfy only the weak hypothesis. For sake of simplicity and because almost all interesting cases are covered, we here restrict ourselves to the situation where the strong hypothesis holds.

\vspace{.2cm}

The next proposition establishes a link between the spectral properties of $T_1^E$ and $H$.

%%%%%%%%%%%%%%%%%%%%%%%%%%%%%%%%%%%%%%%%%%%%%%%
\begin{proposi}
\label{prop-transferopspec} One has:

\begin{enumerate}[\rm (i)] 

\item $E\in \sigma(H_1(k_2))$ for some $k_{2}$ $\Longleftrightarrow$ $e^{\imath \theta}\in \sigma(T_1^E)$ for some $\theta$

\item $E\in \sigma(H)$ $\Longleftrightarrow$ $\sigma(T_1^E)\cap \SM^1\not =\emptyset$
\end{enumerate}
\end{proposi}
%%%%%%%%%%%%%%%%%%%%%%%%%%%%%%%%%%%%%%%%%%%%%%%

\noindent {\bf Proof.} (i) We only prove the implication $"\Longleftarrow"$ because the other is obtained by the inverse procedure. Moreover, let us suppose $p=1$. Let $e^{\imath \theta}$ belong to the spectrum $\sigma(T_1^E)$. Thanks to \eqref{eq-transferFourier}, this implies that there is some $k_{2}\in[-\pi,\pi)$ such that $e^{\imath \theta}\in\sigma(T_1^E(k_{2}))$. For such a $k_{2}$, let $v(k_{2})=(v_0(k_{2}),v_1(k_{2}))\in\CM^{L}\oplus\CM^{L}$ be the corresponding eigenvector, that is
$$
T_1^E(k_{2})v(k_{2})\;=\;e^{\imath \theta}\,v(k_{2})
\;.
$$
The second line of this equation is $A(k_2)^{-1}v_0(k_2)=e^{\imath \theta} v_1(k_2)$.  As $A(k_2)^{-1}$ has no kernel, this shows that neither $v_0(k_2)$ nor $v_1(k_2)$ are vanishing. The first line then becomes
$$
Ev_1(k_2)\;=\;
e^{\imath \theta} A(k_2)v_1(k_2)+ B(k_2)v_1(k_2)+
e^{-\imath \theta} A(k_2)^*v_1(k_2)
\;.
$$
Therefore $\phi\in\ell^\infty(\ZM)\otimes\CM^{L}$ defined by $\phi(n_1)=e^{\imath \theta n_1}v_1(k_2)$ satisfies $H_1(k_2)\phi=E\phi$.  From this we can now readily construct a Weyl sequence for $H_1(k_2)$ at energy $E$. Let $\chi_N\in\ell^2(\ZM)\otimes\CM^{L}$ be the indicator function to $ [-N,N]$. Then $\|\chi_N\phi\|=\Oo(N^{\frac{1}{2}})$ and  set  $\phi_N=\chi_N\phi/\|\chi_N\phi\|$. It follows that  $\|(H_1(k_2)-E)\phi_N\|=\Oo(N^{-\frac{1}{2}})$ and we conclude that $E\in\sigma(H_1(k_2))$. Let us note that by translating the $\phi_N$ one can also obtain an orthonormal Weyl sequence so that $E$ is actually in the essential spectrum of $H_1(k_2)$.

\vspace{.2cm}

\noindent (ii) Follows directly from (i), \eqref{eq-transferFourier} and the analog of \eqref{eq-Hfibered}.
\hfill $\Box$

\vspace{.2cm}

The proposition implies that the transfer operator $T_1^E$ is in $\GUM(\Kk,I)$ if and only if $E\not\in\sigma(H)$. As they do not have any spectrum on $\SM^1$, all the invariants introduced in Section~\ref{app} are necessarily trivial though.

%%%%%%%%%%%%%%%%%%%%%%%%%%%%%%%%%%%%%%%%%%%%%%%%
\section{Transfer operators of the half-plane models}
\label{sec-transferhalf}

The half-plane transfer matrices $\widehat{T}_1^E$ are defined by the same formula \eqref{eq-transferopdef} as $T_1^E$, but with $A$ and $B$ replaced by $\widehat{A}$ and $\widehat{B}$, or differently stated, with $S_2$ replaced by $\widehat{S}_2$. The operator $\widehat{T}_1^E$ acts on the Krein space $\widehat{\Kk}=\ell^2(\NM)\otimes\CM^L\otimes\CM^2$ with the fundamental symmetry $I$ acting on the fiber $\CM^2$. To assure boundedness of $\widehat{T}_1^E$, we we will make the following assumption that again holds in most concrete situations.

\vspace{.2cm}

\noindent {\bf Half-space Strong Invertibility Hypothesis:} {\sl $\widehat{A}$ is invertible with bounded inverse.} 

\vspace{.2cm}

\noindent Again it is possible to write out the solution of the Schr\"odinger equation $\widehat{H}\psi=E\psi$ with the transfer operators, similar as in \eqref{eq-transferrel}.  It is, however, not possible to diagonalize $\widehat{T}_1^E$ by Fourier transform in order to determine the spectrum of $\widehat{T}_1^E$. Using the Weyl sequences of the proof of Proposition~\ref{prop-transferopspec} (which can be shifted away from the boundary), one can readily check that the spectra satisfy $\sigma(T_1^E)\subset\sigma(\widehat{T}_1^E)$. Furthermore, $\Kk=\widehat{\Kk}\oplus\widehat{\Kk}$ and, if $\widetilde{T}^E_1$ is the opposite half-plane transfer matrix defined by replacing $S_2$ in $T_1^E$ by the partial isometry $\widetilde{S}_2$ defined by $(\widetilde{S}_2)^*\widetilde{S}_2=\one$ and $\widetilde{S}_2(\widetilde{S}_2)^*=\one-|0\rangle\langle 0|$, then
$$
T_1^E\;=\;\widehat{T}_1^E\oplus\widetilde{T}^E_1
\,+\,
K\;,
$$ 
where $K$ is a finite range operator which can easily be written out explicitly. The spectrum of $\widetilde{T}^E_1$ and $\widehat{T}_1^E$ are expected to be of similar nature (but not identical), and hence of same nature as the spectrum of the compact perturbation $T_1^E-K$ of $T_1^E$.   For self-adjoint operators, Weyl's theorem tells us that the essential spectra (all but the discrete spectrum) of a given self-adjoint operator and its compact perturbation coincide. For general non-self-adjoint operators, analytic Fredholm theory applies. It shows that the components of the resolvent set of $T_1^E$ either only contain discrete spectrum of $T_1^E-K$ or become point spectrum of $T_1^E-K$ (see \cite{SB} for a detailed proof of this fact). This second possibility is well-known to appear when the two-sided shift $S$ (with spectrum $\SM^1$) is cut into a one-sided shift $\widehat{S}$ and its adjoint $\widehat{S}^*$ (with spectrum given by the filled unit disc). This may also happen within the class of $J$-unitaries on a Krein space, as illustrated by example in \cite[Section 6.4]{SB}. Actually, in the present context this second possibility is connected with the appearance of a flat band of edge states, as  will be argued below based on the following proposition.

%%%%%%%%%%%%%%%%%%%%%%%%%%%%%%%%%%%%%%%%%%%%%%%
\begin{proposi}
\label{prop-transferophatpointsectrum}
One has
$$
\mbox{\rm geometric mult. of $e^{\imath k_1}$ as eigenvalue of }\widehat{T}_1^E
\;=\;
\mbox{\rm mult. of $E$ as eigenvalue of }\widehat{H}_2(k_1)
\;.
$$
\end{proposi}
%%%%%%%%%%%%%%%%%%%%%%%%%%%%%%%%%%%%%%%%%%%%%%%

\noindent {\bf Proof.} "$\geq$" Let $\phi\in\ell^2(\NM)\otimes\CM^{L}$ be such that $\widehat{H}_2(k_1)\phi=E\phi$. Define $\psi$ by $\psi(n_1,n_2)=e^{\imath k_1n_1}\phi(n_2)$, which is not in the Hilbert space $\widehat{\Hh}=\ell^2(\ZM\otimes\NM)\otimes \CM^L$. One checks that $\widehat{H}\psi=E\psi$ so that $\psi$ satisfies the half-plane version of \eqref{eq-transferrel}. Therefore replacing $\psi$ gives
\begin{equation}
\label{eq-Mhateigenvalue}
e^{\imath k_1}
\,\begin{pmatrix}
\widehat{A}\, e^{\imath k_1}\phi \\ \phi
\end{pmatrix}
\;=\;
\widehat{T}_1^E\,
\begin{pmatrix}
\widehat{A}\,e^{\imath k_1}\phi \\ \phi
\end{pmatrix}\;,
\end{equation}
showing that $e^{\imath k_1}$ is an eigenvalue of $\widehat{T}_1^E$. Furthermore, an orthonormal family of eigenvectors of $\widehat{H}_2(k_1)$ with eigenvalue $E$  leads to a family of linearly independent eigenvectors of $\widehat{T}_1^E$.

\vspace{.2cm}

\noindent "$\leq$" Let $\phi,\phi'\in\ell^2(\NM)\otimes\CM^{L}$ be such that
$$
\widehat{T}_1^E\,
\begin{pmatrix}
\phi' \\ \phi
\end{pmatrix}
\;=\;
e^{\imath k_1}
\,\begin{pmatrix}
\phi' \\ \phi
\end{pmatrix}
\;.
$$
Due to the form \eqref{eq-transferopdef}, the second line of this equation is $\widehat{A}^{-1}\phi'=e^{\imath k_1}\phi$ which implies $\phi'=e^{\imath k_1}\widehat{A}\phi$.  In particular, $\phi$ does not vanish. Now again \eqref{eq-Mhateigenvalue} holds and therefore $\psi(n_1,n_2)=e^{\imath k_1n_1}\phi(n_2)$ is a generalized eigenvector of $\widehat{H}$ to the eigenvalue $E$. Taking the partial Fourier transform $\Uu_1$ then shows that $\phi$ is an eigenvector of $\widehat{H}_2(k_1)$ with eigenvalue $E$. Again linearly independent eigenvectors of $\widehat{T}_1^E$ produce linearly independent eigenvectors of $\widehat{H}_2(k_1)$.
\hfill $\Box$

\vspace{.2cm}

Combined with the analytic Fredholm theory, this implies the following result.

%%%%%%%%%%%%%%%%%%%%%%%%%%%%%%%%%%%%%%%%%%%%%%%
\begin{coro}
\label{coro-transferophalfsectrum}
Let $E\not\in\sigma(H)$. Then one of the following holds:

\begin{enumerate}[\rm (i)] 

\item The spectrum of $\widehat{T}_1^E$ has only a finite number of discrete eigenvalues on $\SM^1$ which give rise to  a finite number of edge bands at $E$. 

\item All points on $\SM^1$ are eigenvalues of $\widehat{T}_1^E$ giving rise to a flat band of edge states at $E$.

\end{enumerate}
\end{coro}
%%%%%%%%%%%%%%%%%%%%%%%%%%%%%%%%%%%%%%%%%%%%%%%

Item (i) states that $\widehat{T}_1^E$ is an element of the set $\GUM(\widehat{\Kk},I)$ of essentially $\SM^1$-gapped $I$-unitaries, as more formally defined in Section~\ref{app}. An example of case (ii) is discussed in \cite{NGK,GP}. In this case, $\widehat{T}_1^E$ lies in the class of $\SM^1$-Fredholm operators discussed in \cite{SB,SV}, but is not essentially $\SM^1$-gapped. On the other hand, in case (i) the property $\widehat{T}_1^E\in\GUM(\widehat{\Kk},I)$ allows to define $\Sig(\widehat{T}_1^E)$ which may take any value in $\ZM$. Hence the signature allows to distinguish transfer operators of different systems. It was already shown in \cite{SB} that the transfer operators associated to the Harper model at various energies provide examples of operators in $\GUM(\widehat{\Kk},I)$ with different values of the signature. If $\Sig(\widehat{T}_1^E)=0$, then by a succession of Krein collisions $\widehat{T}_1^E$ can be continuously deformed within $\GUM(\widehat{\Kk},I)$ into an operator with no spectrum on the unit circle \cite{SV}. Whether this path can be constructed by a homotopy of the Hamiltonian has to be analyzed in a given concrete situation. If this is possible, then this homotopy leads to a Hamiltonian with no edge spectrum at $E$. Such a homotopy certainly does not exist if $\Sig(\widehat{T}_1^E)\not =0$. 
In the next Subsection~\ref{sec-calcsig}, the calculation of the edge states is recapitulated from \cite{ASV} and is reinterpreted as a technique to calculate the unit eigenvalues of the half-sided transfer operators (thereby fully developing the ideas only sketched in Section~6.4 of \cite{SB}). In the following Section~\ref{sec-calcsig2} it is shown how to calculate their Krein inertia. Then in Sections~\ref{sec-TRIop} and \ref{sec-PHop} it is shown how symmetries of the Hamiltonian lead to Real symmetries of the transfer operators.

%%%%%%%%%%%%%%%%%%%%%%%%%%%%%%%%%%%%%%%%%%%%%%%%
\subsection{Calculation of the eigenvalues on $\SM^1$}
\label{sec-calcsig}

The calculation of the spectrum of $\widehat{T}_1^E$ on the unit circle will be based on Proposition~\ref{prop-transferophatpointsectrum}. Its proof shows that for $\phi\in\ell^2(\NM)\otimes\CM^L$ the following equivalence holds:
\begin{equation}
\label{eq-basiccalcrel}
\widehat{H}_2(k_1)\phi\;=\;E\phi
\qquad
\Longleftrightarrow
\qquad
\widehat{T}_1^E\,
\begin{pmatrix}
\widehat{A}\,e^{\imath k_1}\phi \\ \phi
\end{pmatrix}\;=\;
e^{\imath k_1}
\,\begin{pmatrix}
\widehat{A}\, e^{\imath k_1}\phi \\ \phi
\end{pmatrix}
\;,
\end{equation}
and that every eigenvector of $\widehat{T}_1^E$ is of the form given on the r.h.s.. Hence the spectrum of $\widehat{T}_1^E$ can be calculated by determining all $k_1$ for which $E$ is eigenvalue of $\widehat{H}_2(k_1)$. This can be done by intersection theory as explained in the appendix of \cite{ASV} and briefly recapitulated here (see also \cite{AEG}). First let us rewrite \eqref{eq-H2k1} as
$$
\widehat{H}_2(k_1)
\;=\;
\widehat{S}_2^*\,a(k_1)\,+\,b(k_1)\,+a(k_1)^*\,\widehat{S}_2
\;,
$$
where
\begin{align*}
& a(k_1)
\;=\;
W_4^*\,e^{\imath(\varphi X_2-k_1)}\,+\,W_2\,+\,W_3^*\,e^{-\imath(\varphi X_2-k_1)}
\;,
\\
& b(k_1)\;=\;
W_1\,e^{\imath(\varphi X_2-k_1)}\,+\,V\,+\,W_1^*\,e^{-\imath(\varphi X_2-k_1)}
\;.
\end{align*}
Both $a(k_1)$ and $b(k_1)$ are matrix-valued multiplication operators on $\ell^2(\NM)\otimes\CM^L$ specified by sequences $(a_n(k_1))_{n\in\NM}$ and $(b_n(k_1))_{n\in\NM}$ of $L\times L$ matrices. Furthermore, if $\varphi=2\pi\frac{q}{p}$, these sequences are $p$-periodic. Let us point out that $a(k_1)$ only commutes with $S_2$ and $S_2^*$ if $\varphi=0$. Now one can introduce the transfer matrices in the $2$-direction at site $n$ and energy $E$ by
$$
T_{2,n}^E(k_1)
\;=\;
\begin{pmatrix}
(E-b_n(k_1))a_n(k_1)^{-1} & -a_n(k_1)^* \\ a_n(k_1)^{-1} & 0
\end{pmatrix}
\;,
$$
whenever the $L\times L$ matrix $a_n(k_1)$ is invertible. These transfer matrices are the fibers of the Fourier transform $\Uu_1$ of the transfer operator $T^E_2$ in the $2$-direction, thus acting on $\ell^2(\ZM)\otimes\CM^L\otimes\CM^2$. If $\varphi=2\pi\frac{q}{p}$, they satisfy the $p$-periodicity relation $T_{2,n+p}^E=T_{2,n}^E$ and similarly for $T^E_2(k_1)$. Again the transfer matrices $T^E_2(k_1)$ allow to write out the formal solutions of the Schr\"odinger equation $H_2(k_1)\phi=E\phi$ and the aim is now to determine the square-integrable solutions which lead to eigenvalues of $H_2(k_1)$. Due to the $p$-periodicity it is therefore natural to introduce the transfer matrix over one $p$-periodicity cell which we choose to be the first $p$ points:
$$
T_{2,p,1}^E(k_1)
\;=\;
T_{2,p}^E(k_1)\cdots
T_{2,1}^E(k_1)
\;.
$$
If the Dirichlet boundary condition (given by an $I$-Lagrangian plane) has a non-trivial intersection with the contracting directions of $T_{2,p,1}^E(k_1)$, then $E$ is an eigenvalue. Now $E$ is chosen to lie in a gap of the spectrum of $H$. Hence the generalized eigenspaces of $T_{2,p,1}^E(k_1)$ associated with all eigenvalues of modulus strictly less than $1$ (namely the contracting ones) constitute an $L$-dimensional $I$-Lagrangian plane in $\CM^{2L}$. Let a basis of this space form the column vectors of a $2L\times L$ matrix $\Phi^E(k_1)$ and then define a unitary $L\times L$ matrix $U^E(k_1)$ by
\begin{equation}
\label{eq-UPhiCalc}
U^E(k_1)\;=\;\binom{\one}{\imath\,\one}^*\Phi^E(k_1)\left( \binom{\one}{-\imath\,\one}^*\Phi^E(k_1)\right)^{-1}
\;.
\end{equation}
Intersection theory now says \cite{ASV} that the dimension of the intersection of the Dirichlet condition with $\Phi^E(k_1)$ is equal to the multiplicity of $1$ as eigenvalue of $U^E(k_1)$. As by \eqref{eq-basiccalcrel} the dimension of this intersection is also equal to the multiplicity of $E$ as eigenvalue of $H_2(k_1)$, this provides a procedure for the calculation of the eigenvalue of $T^E_1$. This can easily be implemented numerically in concrete situations, see Section~\ref{sec-examples}.

%%%%%%%%%%%%%%%%%%%%%%%%%%%%%%%%%%%%%%%%%%%%%%%%
\subsection{Calculation of the signature of eigenvalues on $\SM^1$}
\label{sec-calcsig2}

In the last Subsection~\ref{sec-calcsig} it was explained how a unit eigenvalue $e^{\imath k_1}$ of $\widehat{T}^E_1$ can be calculated based on \eqref{eq-basiccalcrel}. The next aim is to provide tools for the calculation of the inertia $\nu_\pm(e^{\imath k_1})$ which  are needed for the calculation of the various homotopy invariants. The basic fact allowing to do this is that the inertia determine the sign of the rotation of the unit eigenvalue as a function of energy $E$. In order to simplify the calculations let us restrict to the generic case of a simple eigenvalue. 

%%%%%%%%%%%%%%%%%%%%%%%%%%%%%%%%%%%%%%%%%%%%%%%
\begin{proposi}
\label{prop-eigenphasederivative}
Let $e^{\imath k_1(E)}$ be a simple unit eigenvalue of $\widehat{T}^E_1$ with eigenvector $\psi_E$. Then
\begin{equation}
\label{eq-kEderiv}
\partial_E k_1(E)
\;=\;
\frac{\psi_E^*\,(\widehat{T}^E_1)^*\,I\,\partial_E\widehat{T}^E_1\,\psi_E}{\psi_E^*\,\imath\,I\,\psi_E}
\;.
\end{equation}
Furthermore, $\psi_E^*\,(\widehat{T}^E_1)^*I\partial_E\widehat{T}^E_1\psi_E>0$. 
\end{proposi}
%%%%%%%%%%%%%%%%%%%%%%%%%%%%%%%%%%%%%%%%%%%%%%%

\noindent {\bf Proof.}  As $\widehat{T}^E_1$ depends analytically on $E$, so does the simple eigenvalue and its eigenvector, at least in a small interval around $E$.  Deriving the eigenvalue equation $\widehat{T}^E_1\psi_E=e^{\imath k_1(E)}\psi_E$, one finds
$$
(\partial_E\widehat{T}^E_1)\psi_E\,+\,\widehat{T}^E_1\partial_E\psi_E
\;=\;
\imath\,\partial_E k_1(E)\,e^{\imath k_1(E)}\psi_E
\,+\,
e^{\imath k_1(E)}\,\partial_E\psi_E
\;.
$$
Now let us multiply this equation from the left by $(\psi_E)^*I$. As $(\psi_E)^*I\widehat{T}^E_1=e^{\imath k_1(E)}(\psi_E)^*I$, it follows that
$$
(\psi_E)^*\,I\,(\partial_E\widehat{T}^E_1)\,\psi_E
\;=\;
\imath\,\partial_E k_1(E)\;(\psi_E)^*\,I\,\psi_E
\;.
$$
This leads to the formula for $\partial_E k_1(E)$. Furthermore,
$$
(\widehat{T}^E_1)^*\,I\,\partial_E\widehat{T}^E_1
\,=
\begin{pmatrix}
(\widehat{A}^{-1})^*(E-\widehat{B}) & (\widehat{A}^{-1})^* \\
- \widehat{A} & 0 
\end{pmatrix}
\begin{pmatrix}
0 & -\one \\
\one & 0 
\end{pmatrix}
\begin{pmatrix}
\widehat{A}^{-1} & 0 \\
0 & 0 
\end{pmatrix}
=
\begin{pmatrix}
(\widehat{A}^{-1})^*\widehat{A}^{-1} & 0 \\
 0 & 0
\end{pmatrix}
.
$$
As the eigenvectors of $\widehat{T}^E_1$ never have a vanishing upper component, this shows the claimed positivity and therefore concludes the proof. 
\hfill $\Box$

\vspace{.2cm}

As the sign of $\psi_E^*\,\imath \,I\psi_E$ is by definition the Krein inertia $(\nu_+(\lambda),\nu_-(\lambda))$ of $\lambda=e^{\imath k_1(E)}$ (see Section~\ref{app} for the definition), Proposition~\ref{prop-eigenphasederivative} implies the following.

%%%%%%%%%%%%%%%%%%%%%%%%%%%%%%%%%%%%%%%%%%%%%%%
\begin{coro}
\label{coro-eigenphasederivative}
Let $e^{\imath k_1(E)}$ be a simple unit eigenvalue of $\widehat{T}^E_1$ with eigenvector $\psi_E$. Then
\begin{equation}
\label{eq-kEderivsign}
\mbox{\rm sign}\bigl(\partial_E k_1(E)\bigr)
\;=\;
\nu_+(e^{\imath k_1(E)})\,-\,\nu_-(e^{\imath k_1(E)})
\;.
\end{equation}
\end{coro}
%%%%%%%%%%%%%%%%%%%%%%%%%%%%%%%%%%%%%%%%%%%%%%%

Now let $E(k_1)$ denote the eigenvalue of $\widehat{H}_2(k_1)$ in \eqref{eq-basiccalcrel}. Then one has $\partial_E k_1(E)=\bigl(\partial_{k_1}E(k_1)\bigr)^{-1}$. Furthermore, by Proposition~1 of \cite{ASV}, 
$$
\partial_{k_1}E(k_1)
\;=\;
\frac{1}{2}\;\partial_{k_1}\theta^E(k_1)
\;,
$$
where $e^{\imath \theta^E(k_1)}$ is the eigenvalue of $U^E(k_1)$ with $e^{\imath \theta^E(k_1)}=1$ giving rise to the eigenvalue of $\widehat{H}_2(k_1)$ by the intersection theory described in Section~\ref{sec-calcsig}. Therefore, one has the following fact that allows to calculate the signature by analyzing the finite-dimensional matrices $U^E(k_1)$.

%%%%%%%%%%%%%%%%%%%%%%%%%%%%%%%%%%%%%%%%%%%%%%%
\begin{coro}
\label{coro-eigenphasederivative2}
Let $e^{\imath k_1(E)}$ be a simple unit eigenvalue of $\widehat{T}^E_1$. Then its signature is given by
$$
\nu_+(e^{\imath k_1(E)})\,-\,\nu_-(e^{\imath k_1(E)})
\;=\;
\mbox{\rm sign}\bigl(
\partial_{k_1}\theta^E(k_1)\bigr)
\;,
$$
where $e^{\imath \theta^E(k_1)}$ is the eigenvalue of $U^E(k_1)$ with $e^{\imath \theta^E(k_1)}=1$.
\end{coro}
%%%%%%%%%%%%%%%%%%%%%%%%%%%%%%%%%%%%%%%%%%%%%%%

As in applications to concrete models (Section~\ref{sec-examples}) it is simple to plot the spectrum of $U^E(k_1)$ as a function of $k_1$, Corollary~\ref{coro-eigenphasederivative2} allows to determine the Krein inertia of $\widehat{T}^E_1$ at least for simple eigenvalues.

\vspace{.2cm}

From the Krein inertia of all eigenvalues on the unit circle one can deduce $\Sig(\widehat{T}^E_1)$ by \eqref{eq-sigdef} which is then a homotopy invariant. In particular, it does not change as $E$ varies as long as the gap of $H$ remains open or equivalently $\widehat{T}^E_1$ remains essentially $\SM^1$-gapped. Due to Corollary~\ref{coro-eigenphasederivative2}, one readily checks that  $\Sig(\widehat{T}^E_1)$ is equal to the edge index $\mbox{Ei}(E)$ defined in \cite{ASV}. Furthermore, this invariant is simply connected to the Chern number $\Ch(P_E)$ of the Fermi projection $P_E=\chi(H\leq E)$ of the planar Hamiltonian. The reader is referred to \cite{KRS,ASV} for the definition of the Chern number as well as the proof of the following result.

%%%%%%%%%%%%%%%%%%%%%%%%%%%%%%%%%%%%%%%%%%%%%%%
\begin{theo}
\label{theo-SigChern}
Let $E$ lie in a gap of $H$. Then $\Sig(\widehat{T}^E_1)=\Ch(P_E)$.
\end{theo}
%%%%%%%%%%%%%%%%%%%%%%%%%%%%%%%%%%%%%%%%%%%%%%%

%%%%%%%%%%%%%%%%%%%%%%%%%%%%%%%%%%%%%%%%%%%%%%%%
\section{Random perturbations}
\label{sec-rand}

All the above is based on the study of the transfer operators $T^E_1$ and $\widehat{T}^E_1$. These transfer operators may still contain a random perturbation stemming, {\it e.g.}, from a random potential which is included in $V$ in \eqref{eq-genH}. The system corresponding to these transfer matrices is still periodic in the $1$-direction, even though it lost periodicity in the $2$-direction (similar to the situation studied in \cite{GP}). The random transfer operators are still $I$-unitary and, if the random term is small compared to the bulk energy gap, they are still essentially $\SM^1$-gapped and therefore still have signature invariants $\Sig$, $\Sig_2$ and $\Sec$. Moreover, when the weak random perturbation is added homotopically, these invariants do not change due to the results described in Section~\ref{app} and are therefore given by the invariants of the periodic models. For systems that are $p$-periodic in the $1$-direction, one can still apply the formalism described above to the $p$-fold product of transfer operators. On the other hand, for systems without periodicity in the $1$-direction a new idea is necessary. 

%%%%%%%%%%%%%%%%%%%%%%%%%%%%%%%%%%%%%%%%%%%%%%%%
\section{Time-reversal symmetry as Real structure}
\label{sec-TRIop}

Let us now suppose that the internal degrees of freedom decompose as  $L=rR$ where $r=2s+1$ results from a spin $s\in\NM/2$ of the particles. Let $s^x$ and $s^z$ be real, and  $s^y$ a purely imaginary $r\times r$ matrices representing the Lie algebra $\mbox{\rm su}(2)$. Then set
$$
\IS\;=\;\e^{\imath \pi s^y}
\;.
$$
Hence $\IS$ is a real unitary and the spectral theory of the spins implies that $\IS^2=(-1)^{2s}\one$. The time-reversal operator on the Hilbert spaces $\Hh$ and $\widehat{\Hh}$ is then given by complex conjugation $\Cc$ followed by $\IS$. The Hamiltonian $H$ has time-reversal symmetry (TRS) if
$$
\IS^*\,\overline{H}\,\IS\;=\;H\;,
$$
and similarly for $\widehat{H}$. The TRS is called even for integer spin ($\IS^2=\one$) and odd for half-integer spin ($\IS^2=-\one$). The TRS of $H$ holds if and only if the magnetic flux $\varphi=0$ mod $\pi$ and the matrices defining $H$ by \eqref{eq-genH} satisfy $\IS^*\,\overline{W_i}\,\IS=W_i$ and $\IS^*\,\overline{V}\,\IS=V$ so that
$$
\IS^*\,\overline{A}\,\IS\;=\;A\;,
\qquad
\IS^*\,\overline{B}\,\IS\;=\;B\;.
$$
The TRS is then passed on to the half-space operators $\widehat{H}$. Only Hamiltonians with TRS will be considered in this section. Then the associated transfer operators satisfy
\begin{equation}
\label{eq-transferopTRI}
(\one\otimes\IS)^*\,\overline{T_1^E}\;\one\otimes\IS
\;=\;
T_1^E
\;,
\qquad
(\one\otimes\IS)^*\,\overline{\widehat{T}_1^E}\;\one\otimes\IS
\;=\;
\widehat{T}_1^E
\;.
\end{equation}
Here the $\one$ is in the grading of \eqref{eq-transferopdef} and $\IS$ acts on the spin degree of freedom of the Krein spaces $(\Kk,I\otimes\one)$ and $(\widehat{\Kk},I\otimes\one)$ respectively. Hence $\one\otimes\IS$ induces a Real structure which is even or odd pending on whether the spin is integer or half-integer. Furthermore, $I\otimes\one$ and $\one\otimes\IS$ commute. Hence the Real Krein spaces $\Kk$ and $\widehat{\Kk}$ are of kind $(-1,1,1)$ or $(-1,-1,1)$ pending on whether the TRS is even or odd (see Section~\ref{app} for the definition of a kind of a Real Krein space).  For any real energy, the transfer operators $T_1^E$ and $\widehat{T}_1^E$ are in $\UM(\Kk,I\otimes\one,\one\otimes\IS)$ and $\UM(\widehat{\Kk},I\otimes\one,\one\otimes\IS)$ respectively (see again Section~\ref{app} for the definition of these classes of operators). Furthermore, by Corollary~\ref{coro-transferophalfsectrum} the transfer operators at $E\not\in\sigma(H)$ are in the class $\GUM(\widehat{\Kk},I\otimes\one,\one\otimes\IS)$ of essentially gapped $I\otimes\one$-unitaries with Real symmetry $\one\otimes\IS$. For even TRS there exists a homotopy from $\widehat{T}_1^E$ to a trivial transfer operator without edge spectrum exists (by adding an adequate finite dimensional perturbation as in \cite{SV}). On the other hand, for odd TRS the multiplicity of the edge spectrum is even (by Kramers degeneracy  as in Proposition~5 of \cite{SV}). Nevertheless, there may be a non-trivial $\ZM_2$-invariant $\Sig_2(\widehat{T}_1^E)$ by \eqref{eq-Sig2Def}. An example for the latter is the Kane-Mele model \cite{KM}, see Section~\ref{sec-KM}. The corresponding physical effect is the quantum spin Hall effect (QSHE) \cite{KM}, see Table~1.

%%%%%%%%%%%%%%%%%%%%%%%%%%%%%%%%%%%%%%%%%%%%%%%%
\section{Particle-hole symmetry as Real structure}
\label{sec-PHop}

The particle-hole symmetry (PHS) is a property of Bogoliubov-de Gennes Hamiltonians describing particles and anti-particles in a superconductor by an effective one-particle Hamiltonian $H$ on $\ell^2(\ZM^2)\otimes\CM^L$ of the form \eqref{eq-genH}. This symmetry is of the form
\begin{equation}
\label{eq-PHS}
\KPH^*\,\overline{H}\,\KPH
\;=\;
-\,H
\;,
\end{equation}
where $\KPH$ is a real unitary operator satisfying $\KPH^2=\etaPH\one$ with $\etaPH=\pm1$. The PHS is then called even and odd if $\etaPH=1$ and $\etaPH=-1$ respectively, and the Hamiltonian with PHS is then said to be of Class D and C respectively \cite{AZ}. The reader is referred to \cite{AZ,SRFL,DDS} for further details on these operators and their physical motivation. Only one further point will be of importance here, namely $H$ depends on a chemical potential $\mu$ which plays the role of an energy parameter (see Sections~\ref{sec-pp} and \ref{sec-dd} for examples). Therefore one is solely interested in zero energy solutions $H\psi=0$ and hence the transfer operator \eqref{eq-transferopdef} somewhat simplifies and will be denoted by $T^\mu_1$ to indicate the dependence on the chemical potential (which is part of the coefficient operator  $B$). Now the coefficient operators in \eqref{eq-HamJac1} of a Hamiltonian with PHS satisfy
$$
\KPH^*\,\overline{A}\,\KPH
\;=\;
-\,A\;,
\qquad
\KPH^*\,\overline{B}\,\KPH
\;=\;
-\,B
\;,
$$
and therefore the transfer operator $T_1^\mu$ satisfies
$$
(J\otimes\KPH)^*
\,\overline{T_1^\mu}\,
J\otimes\KPH
\;=\;T_1^\mu
\;,
$$
where $J=\binom{\one\;\;0}{0\;-\one}$ in the grading of \eqref{eq-transferopfundsym}. This is again a Real symmetry of the $I\otimes \one$-unitary operator $T_1^\mu$ in the Krein space $(\Kk,I\otimes \one)$ with Real symmetry $J\otimes \KPH$. The kind of this Krein space is $(-1,1,-1)$ for even PHS and $(-1,-1,-1)$ for odd PHS (see Section~\ref{app} for the definition of a kind of a Real Krein space). All the above also holds (and is more interesting) for the half-space restriction $\widehat{T}_1^\mu$ of $T_1^\mu$ which acts on the Krein space $\widehat{\Kk}$. According to Table~1 there are respectively $\ZM\times\ZM_2$ and $\ZM$-valued invariants associated to operators in $\GUM(\Kk,I\otimes\one,J\otimes \KPH)$ and $\GUM(\widehat{\Kk},I\otimes\one,J\otimes \KPH)$ in the cases of even and odd PHS respectively. Again the invariants of $T^\mu_1$ are trivial, but those of $\widehat{T}_1^\mu$ allow to distinguish different phases of the thermal quantum Hall effect (TQHE) and the spin quantum Hall effect (SQHE) respectively \cite{SRFL}. Non-trivial examples are given in Sections~\ref{sec-pp} and \ref{sec-dd}.

\vspace{.2cm}

Finally let us take a closer look at the eigenvalues $\pm 1$ of $\widehat{T}_1^\mu$ in the case of an even PHS (Class D). These eigenvalues, generically simple, are then stable under perturbations (see Section~\ref{app} and \cite{SV}). A sufficient criterion for the existence of one such eigenvalue it that $\Sig(\widehat{T}_1^\mu)=\Ch(P_\mu)$ is odd, but also an even signature and eigenvalues at both $1$ and $-1$ is conceivable, and this happens precisely when the secondary invariant $\Sec$ defined in \eqref{eq-SecDef} is non-trivial. If there is a simple eigenvalue at $\pm 1$, then the corresponding eigenstate $\psi$ of $\widehat{T}_1^\mu$ satisfies $\psi=J\otimes\KPH\overline{\psi}$ (see also Proposition~5 of \cite{SV}). Using \eqref{eq-basiccalcrel}, one then constructs from $\psi$ zero energy edge states $\phi$ (planar waves, not in Hilbert space $\widehat{\Hh}$, satisfying $\widehat{H}\phi=0$), for which $\phi=\pm \KPH\overline{\phi}$. This latter relation means by definition that the edge states are Majorana fermion modes. In Section~\ref{sec-pp} we exhibit a model with one such Majorana edge mode. We are not aware of a model with two counterpropagating Majorana modes ($\Sig$ even and $\Sec=1$).

\begin{figure}
\begin{center}
\includegraphics[width=5cm]{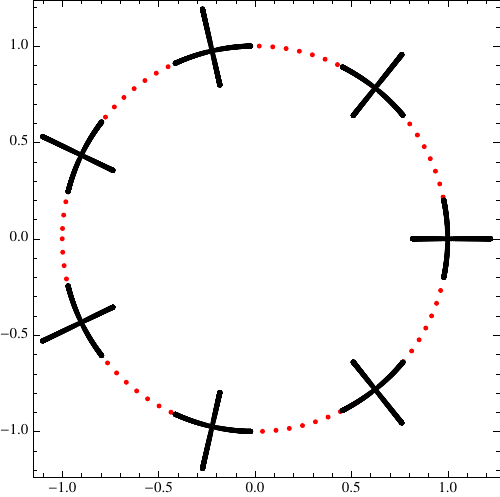}
\hspace{.3cm}
\includegraphics[width=5cm]{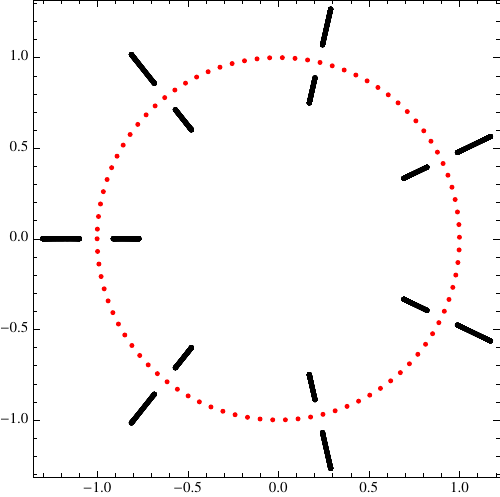}
\hspace{.3cm}
%{\includegraphics[width=4.3cm]{Harper_Flux=,42_E=-1,9.png}}
%\includegraphics[width=4.3cm]{Harper_Flux.pdf}
\includegraphics[width=4.3cm]{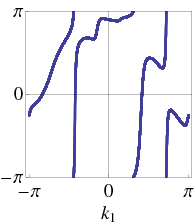}
\caption{\sl For a Harper model with flux $\varphi=2\pi\,\frac{3}{7}$, the spectrum of the transfer operator $T_{1}^E$ is plotted for $E=-2.2$ and $E=-1.9$, together with a dotted unit circle. The energy $E=-2.2$ lies in the spectral band, while $E=-1.9$ lies in a gap. For this latter case, the spectrum of $k_1\in[-\pi,\pi]\mapsto U^E(k_1)$ is plotted in the third figure {\rm (}taken from {\rm \cite{ASV})}. The three intersections $k_1$ of the curves with $0$ provide the positions $e^{\imath k_1}$ of  three eigenvalues of $\widehat{T}^E_1$ on the unit circle. All three are simple and have positive inertia by  {\rm Corollary~\ref{coro-eigenphasederivative2}}.}
\label{fig-Harper}
\end{center}
\end{figure}

%%%%%%%%%%%%%%%%%%%%%%%%%%%%%%%%%%%%%%%%%%%%%%%%
\section{Examples}
\label{sec-examples}

In this final section, some standard physical models with Hamiltonians of the form \eqref{eq-genH} are considered. For each the spectrum of the transfer operator $T^E_1$ is plotted numerically. Its half-space restriction $\widehat{T}^E_1$ then has supplementary discrete spectrum and, by the technique described in Sections~\ref{sec-calcsig} and \ref{sec-calcsig2}, those eigenvalues lying on the unit circle together with their signature can be read off the spectrum of $k_1\mapsto U^E(k_1)$. The eigenvectors of these unit eigenvalues produce edge modes, as explained in Section~\ref{sec-transferhalf}. The non-generic case of a flat band of edge states is not realized by any of the examples below. The first example is the Harper model with rational magnetic which has no Real symmetry. The second is the Kane-Mele model \cite{KM} which has an odd TRS so that the transfer operators $\widehat{T}^E_1$ are of kind $(-1,-1,1)$. Then follow two standard tight-binding BdG models for topological superconductors, namely a $p\pm\imath p$ pair potential leading to transfer operators of kind $(-1,1,-1)$ and a $d\pm\imath d$ pair potential providing transfer operators of kind $(-1,-1,-1)$.  In all these example the invariants collected in Table~1 are non-trivial. All graphics were produced by short mathematica codes with an evaluation taking only a few seconds on a laptop PC.

\vspace{-.2cm}

%%%%%%%%%%%%%%%%%%%%%%%%%%%%%%%%%%%%%%%%%%%%%%%%
\subsection{The Harper model}
\label{sec-Harper}

The Harper model is the magnetic Laplacian on the square lattice with a flux $\varphi=2\pi\,\frac{q}{p}$ through each unit cell. Its Hamiltonian  is $H=U_1+U_1^*+U_2+U_2^*$ where in Landau gauge $U_1=e^{-\imath \varphi X_2}S_1$ and $U_2=S_2$ as above.  In particlular, $L=1$. Then the spectrum of the transfer operator $T^E_1$ given by \eqref{eq-transferopdef} is plotted using the Fourier decomposition \eqref{eq-transferFourier}, for parameter values described in the figure caption. In the left plot there is a band of eigenvalues on the unit circle, so the corresponding energy lies within a band of $H$, while in the middle figure the energy lies in a band gap of $H$ because the transfer operator $T^E_1$ has no spectrum on the unit circle. For the very same energy, the family of unitaries  $k_1\in[-\pi,\pi]\mapsto U^E(k_1)$ is considered and its spectrum is plotted as a function of $k_1$. There are three edge bands with positive phase speed. 

\vspace{-.2cm}

%%%%%%%%%%%%%%%%%%%%%%%%%%%%%%%%%%%%%%%%%%%%%%%%
\subsection{Kane-Mele model}
\label{sec-KM}

The Kane-Mele model \cite{KM} is written out in detail in Section~5.6 of \cite{ASV} and it seems inadequate to repeat the somewhat lengthy description here. It is a model with odd TRS on a honeycomb lattice and with spin $\frac{1}{2}$ so that $L=2\cdot 2=4$. Just as for the Harper Hamiltonian, the spectrum of the transfer operator $T^E_1$ is plotted for one energy in the band of $H$ and for one in the band gap. For the energy in the gap, the third graphic of the figure allows to read off the unit values of $\widehat{T}^E_1$ and their signatures. As there is one eigenvalue on the upper half of the unit circle, the $\ZM_2$-signature is non-trivial. For connections to other $\ZM_2$-invariants as well as physical implications of this non-triviality it is possible to consult \cite{ASV,GS}.

\begin{figure}
\begin{center}
%{\includegraphics[width=5.7cm]{KMTransferE=0,6,LSO=1,LR=0,45,LR=0.png}}
%\includegraphics[width=5.4cm]{KMTransferE=0,6,LSO=1,LR=0,45,LR=0.pdf}
\includegraphics[width=5.4cm]{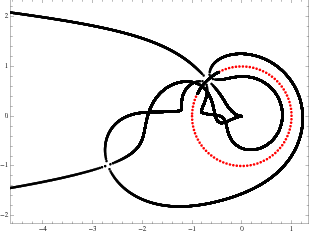}
\hspace{.1cm}
%{\includegraphics[width=6cm]{KMTransferE=0,LSO=1,LR=0,45,LR=0.png}}
%\includegraphics[width=5.7cm]{KMTransferE=0,LSO=1,LR=0,45,LR=0.pdf}
\includegraphics[width=5.7cm]{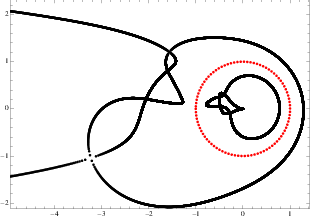}
\hspace{.1cm}
%{\includegraphics[width=4.1cm]{KaneMele_E=0_SO=1_R=,3_ST=4,5.png}}
%\includegraphics[width=3.8cm]{KaneMele_E=0_SO=1_R=,3_ST=4,5.pdf}
\includegraphics[width=3.8cm]{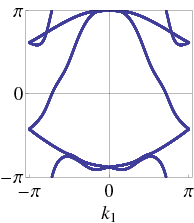}
\caption{\sl For a Kane-Mele model {\rm \cite{KM,ASV}} with parameters $\lambda_{\mbox{\rm\tiny SO}}=1$, $\lambda_{\mbox{\rm\tiny Ra}}=0.45$ and $\lambda_{\mbox{\rm\tiny st}}=0.3$, the spectrum of the transfer operators $T^E_1$ at energies $E=0.6$ and $E=0$ is plotted. For $E=0$ the graphic on the r.h.s. shows the spectrum of $k_1\in[-\pi,\pi]\mapsto U^E(k_1)$. It shows that the $\ZM_2$-invariant of $\widehat{T}^E_1$ for $E=0$ is non-trivial.}
\label{fig-KM}
\end{center}
\end{figure}

\vspace{-.2cm}

%%%%%%%%%%%%%%%%%%%%%%%%%%%%%%%%%%%%%%%%%%%%%%%%
\subsection{BdG model for $p\pm\imath p$ superconductor}
\label{sec-pp}

\begin{figure}
\begin{center}
\includegraphics[width=7cm]{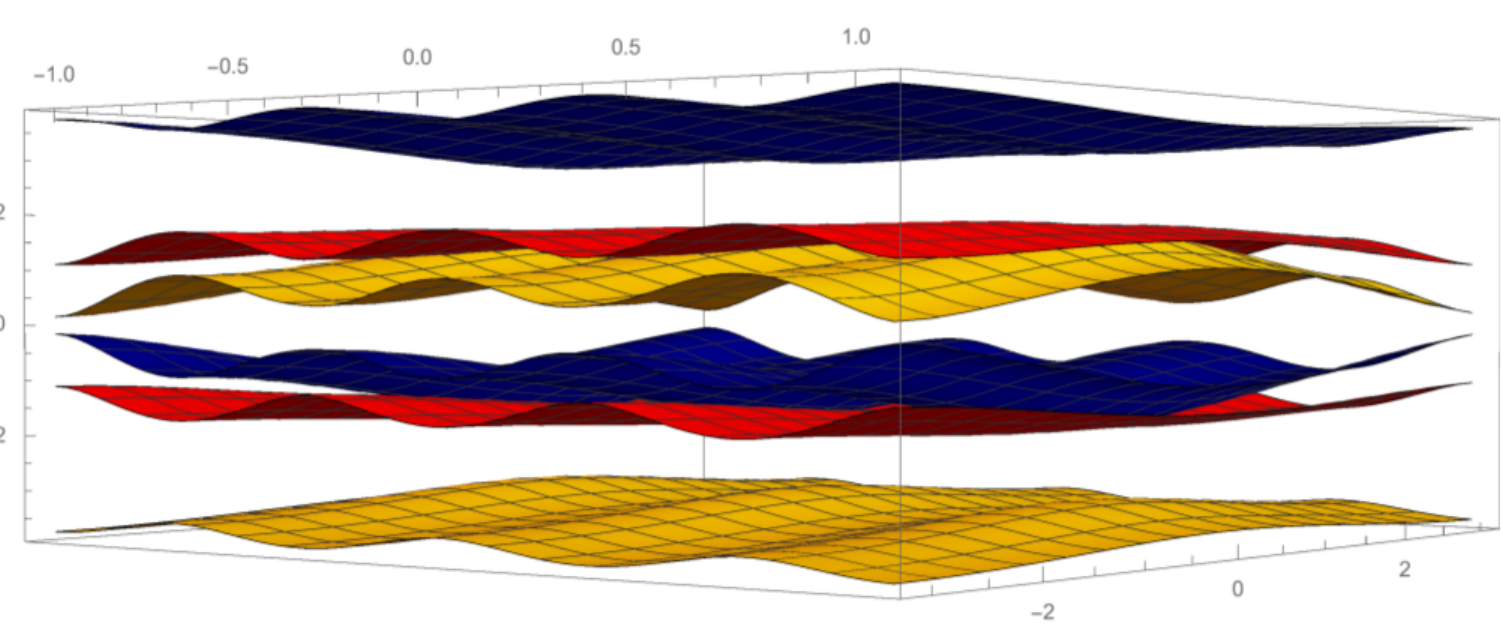}
\hspace{.3cm}
\includegraphics[width=7cm]{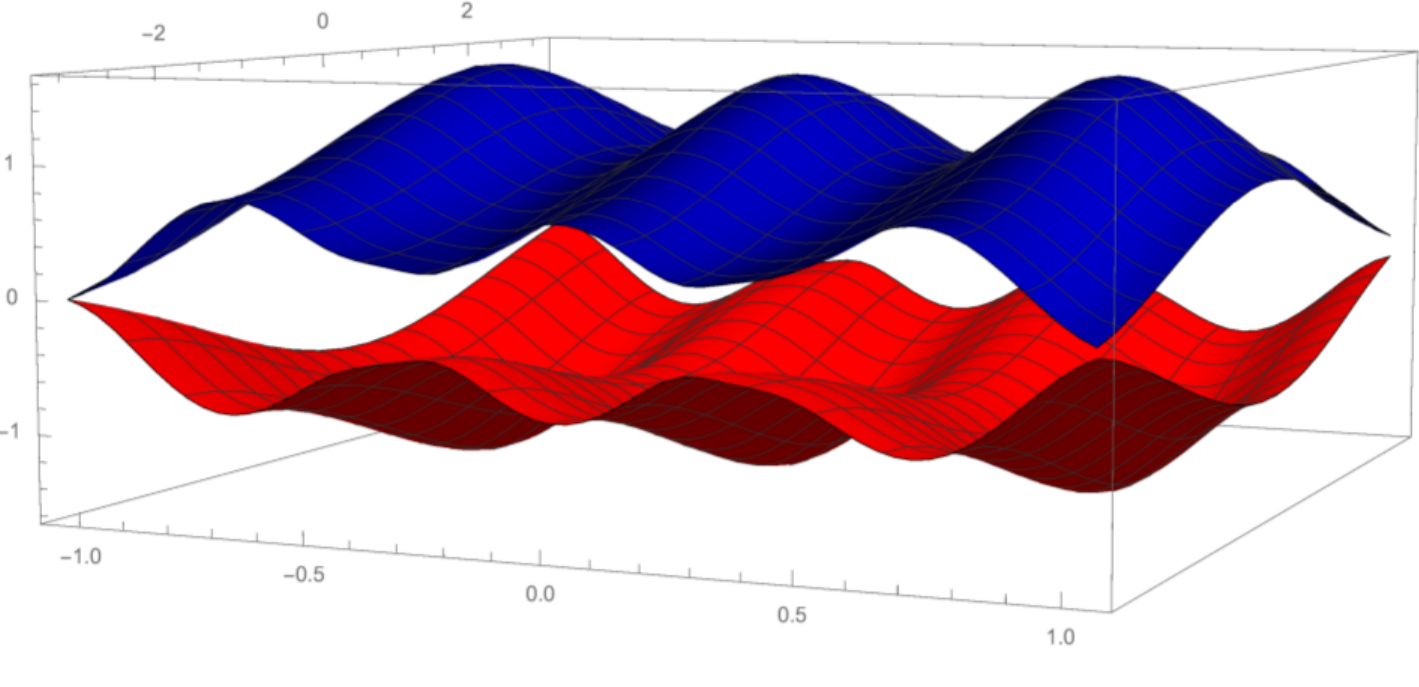}
\caption{\sl This figure shows the energy bands of the $p\pm \imath p$ BdG model with $\delta_p=0.2$ and $\varphi=2\pi\frac{1}{3}$ over the (magnetic) Brillouin zone $[-\frac{2\pi}{3},\frac{2\pi}{3})\times[-\pi,\pi)$. The left plot shows all $6$ bands for $\mu=1.0$. For $\mu=0.2$ the plot looks similar. In particular, in both cases $E=0$ lies in a gap of the spectrum. The right plot shows the $2$ central bands at $\mu=0.813$. They touch at $E=0$ on the corners of the Brillouin zone and produce a pseudo-gap.
}
\label{fig-Bands}
\end{center}
\vspace{-.4cm}
\end{figure}

Also tight-binding BdG Hamiltonians are of the form \eqref{eq-genH}. In the most simple case, the Hilbert space is $\Hh=\ell^2(\ZM^2)\otimes\CM_\ph^2$, hence having only a particle-hole fiber and no other degree of freedom such as spin or sublattice parameter. On this Hilbert space, an $s$-wave superconductor is the easiest, but its edge modes are trivial. Hence we consider a $p\pm\imath p$ pair potential of strength $\delta_p\in\RM$. If $\mu$ denotes the chemical potential, then the corresponding BdG Hamiltonian is given by \cite{SRFL,DDS}
$$
H
\;=\;
%\frac{1}{2}\;
\begin{pmatrix}
U_1+U_1^*+U_2+U_2^* -\mu & \delta_{p}\,(S_1-S_1^*\pm\imath(S_2-S_2^*) )
\\
\delta_{p}\,(S_1^*-S_1\pm\imath(S_2-S_2^*) ) &
-\overline{U}_1-\overline{U}_1^*-\overline{U}_2-\overline{U}_2^*+\mu
\end{pmatrix}
\;.
$$
It can still have a non-trivial magnetic flux $\varphi$ and has the even PHS \eqref{eq-PHS} w.r.t. $S_\ph=\binom{0\;\one}{\one\;0}$. Hence $T^\mu_1$ and $\widehat{T}^\mu_1$ are of kind $(-1,1,-1)$. In Figure~3 is plotted again the spectrum of the transfer operator $T^\mu_1$ for different values of the chemical potential $\mu$. Note that there never appears a band of spectrum on the unit circle, but that only these bands touch the unit circle at a few points. This reflects that Hamiltonian always has a pseudo-gap at energy $0$. Figure~\ref{fig-PPedge} shows the spectrum of $k_1\in[-\pi,\pi]\mapsto U^\mu(k_1)$ for various values of $\mu$ leading to different phases.

\begin{figure}
\begin{center}
\includegraphics[width=5cm]{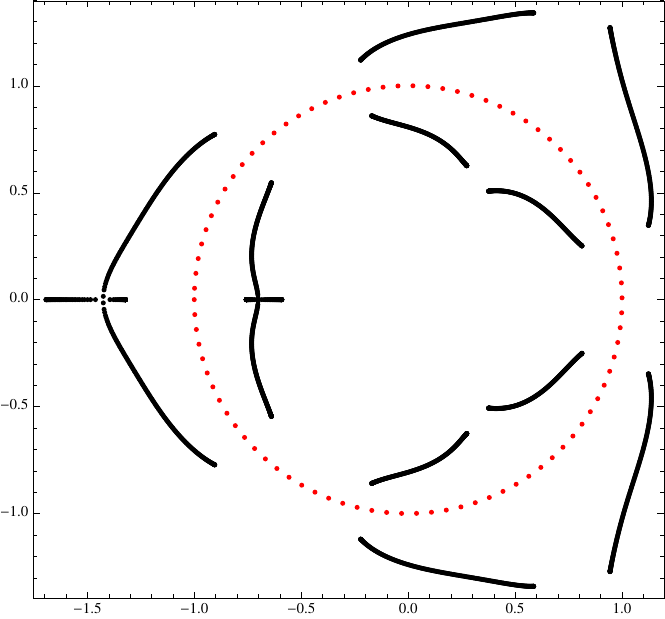}
\hspace{.3cm}
\includegraphics[width=4.8cm]{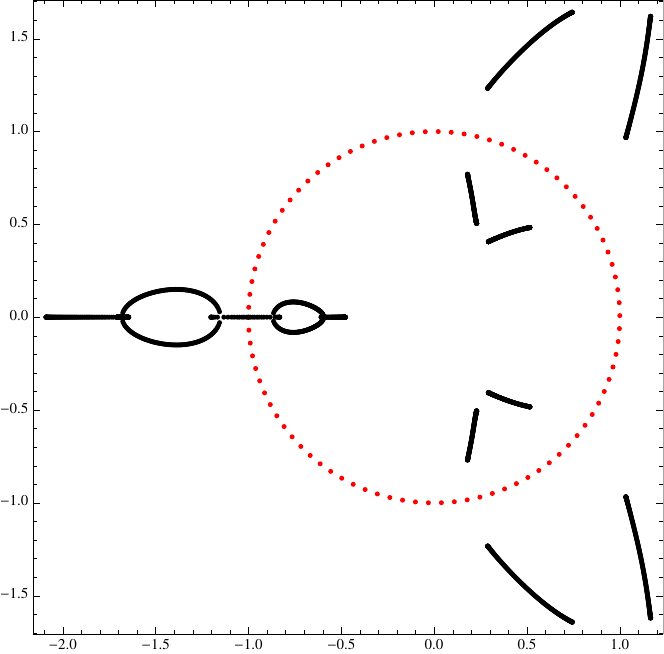}
\hspace{.3cm}
\includegraphics[width=4.8cm]{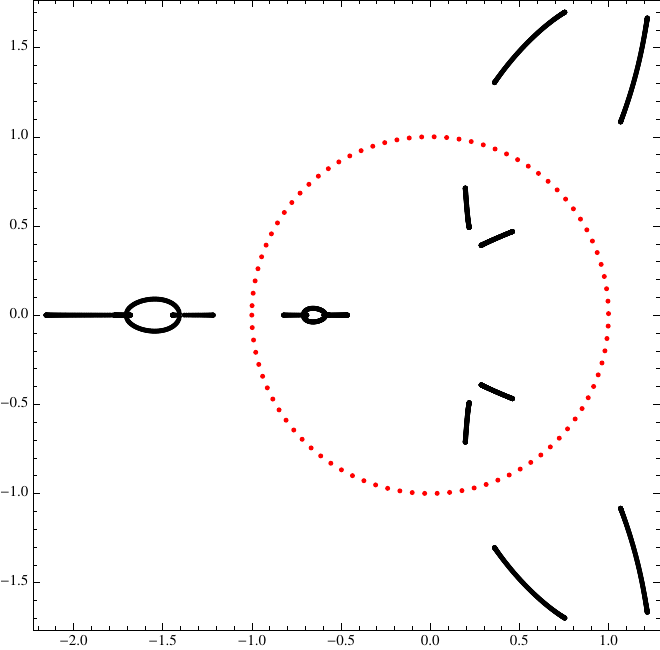}
%\includegraphics[width=5cm]{PPtransferdel0,2mu0,2.pdf}
%\hspace{.3cm}
%\includegraphics[width=4.8cm]{PPtransferdel0,2mu0,813.pdf}
%\hspace{.3cm}
%\includegraphics[width=4.8cm]{PPtransferdel0,2mu1,0.pdf}
\caption{\sl For the $p\pm \imath p$ BdG model with $\delta_p=0.2$ and $\varphi=2\pi\frac{1}{3}$, the spectrum of $T^\mu_1$ is plotted for $\mu=0.2\,,\,0.813\,,\,1.0$. The second figure corresponds to a pseudo-gap.}
\label{fig-PPtransfer}
\end{center}
\vspace{-.4cm}
\end{figure}

\vspace{-.2cm}

%%%%%%%%%%%%%%%%%%%%%%%%%%%%%%%%%%%%%%%%%%%%%%%%
\subsection{BdG model for $d\pm\imath d$ superconductor}
\label{sec-dd}

This section analyzes a BdG Hamiltonian $H$ on $\Hh=\ell^2(\ZM^2)\otimes\CM_\ph^2$ with a $d\pm\imath d$ pair potential of strength $\delta_d\in\RM$ (see \cite{SRFL,DDS}):
$$
%\frac{1}{2}\;
\begin{pmatrix}
U_1+U_1^*+U_2+U_2^* -\mu & \!\!\!\!\!\!\!\!\!\!\!\!\!\!\!\!\!\!\!\!\!\!\! \delta_{d}\,(
S_1+S_1^*-S_2-S_2^*
\pm \imath(S_1-S_1^*)(S_2-S_2^*))
\\
\delta_d(S_1+S_1^*-S_2-S_2^*\mp \imath(S_1-S_1^*)(S_2-S_2^*))
 & \!\!\!\!\!\!\!
-\overline{U}_1-\overline{U}_1^*-\overline{U}_2-\overline{U}_2^*+\mu
\end{pmatrix}
.
$$
It has the odd PHS is implemented by $S_\ph=\binom{0\;-\one}{\one\;\;\;0}$. Therefore the transfer operators $T^\mu_1$ and $\widehat{T}^\mu_1$ are of kind $(-1,-1,-1)$. Again the spectrum of the transfer operators can be plotted as well as the spectrum of $U^\mu(k_1)$. For example, the half-signature for $\varphi=0$, $\mu=0$ and $\delta_d>0$ is $1$. To avoid repetition, no numerical plots are provided.

%%%%%%%%%%%%%%%%%%%%%%%%%%%%%%%%%%%%%%%%%%%%%%%%
\section{Conclusion}
\label{sec-conclusion}

\vspace{-.2cm}

For two-dimensional periodic tight-binding Hamiltonians restricted to a half-space, this paper develops a new technique to address the existence of edge states and to analyze and classify their properties. It is based on a detailed spectral analysis of the infinite-dimensional transfer operators along the boundary of the system. For energies outside of the bulk spectrum, these transfer operators only have isolated eigenvalues on the unit circle. The Krein signatures of these unit eigenvalues determines the sign of the group velocity of the corresponding edge states. The general theory is put work numerically on several models describing two-dimensional topological insulators.

\vspace{.2cm}

\noindent {\bf Acknowledgements:}  We thank PAPPIT-UNAM IN 104015 as well as the DFG SCHU-1356/6 for financial support.

\vspace{-.5cm}

\begin{figure}
\begin{center}
%{\includegraphics[width=5cm]{PPedgedel0,2mu0,2.png}}
%\includegraphics[width=4.7cm]{PPedgedel0,2mu0,2.pdf}
%\hspace{.3cm}
%{\includegraphics[width=5cm]{PPedgedel0,2mu0,813.pdf}}
%\hspace{.3cm}
%{\includegraphics[width=5cm]{PPedgedel0,2mu1,0.pdf}}
%\hspace{.3cm}
%{\includegraphics[width=5cm]{PPedgedel0,2mu1,9.png}}
%\includegraphics[width=4.7cm]{PPedgedel0,2mu1,9.pdf}
%\hspace{.3cm}
%{\includegraphics[width=5cm]{PPedgedel0,2mu2,1.pdf}}
%\hspace{.3cm}
%{\includegraphics[width=5cm]{PPedgedel0,2mu2,5.png}}
%\includegraphics[width=4.7cm]{PPedgedel0,2mu2,5.pdf}
\includegraphics[width=4.7cm]{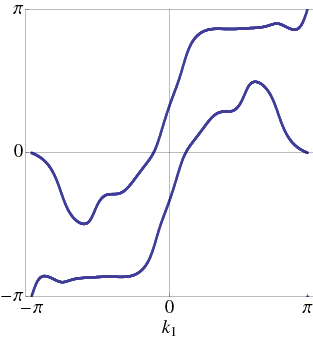}
\hspace{.3cm}
%{\includegraphics[width=5cm]{PPedgedel0,2mu0,813.pdf}}
%\hspace{.3cm}
%{\includegraphics[width=5cm]{PPedgedel0,2mu1,0.pdf}}
%\hspace{.3cm}
%{\includegraphics[width=5cm]{PPedgedel0,2mu1,9.png}}
\includegraphics[width=4.7cm]{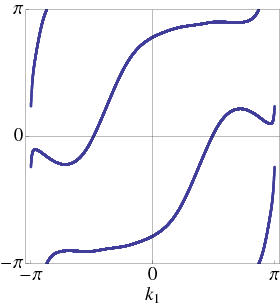}
\hspace{.3cm}
%{\includegraphics[width=5cm]{PPedgedel0,2mu2,1.pdf}}
%\hspace{.3cm}
%{\includegraphics[width=5cm]{PPedgedel0,2mu2,5.png}}
\includegraphics[width=4.7cm]{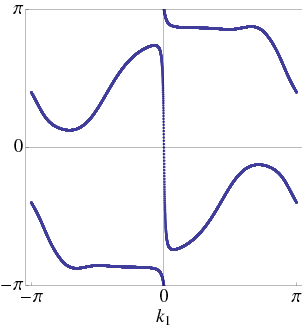}
\caption{\sl For the $p\pm \imath p$ BdG model with $\delta_p=0.2$ and $\varphi=2\pi\frac{1}{3}$, the spectrum of the unitaries $k_1\in[-\pi,\pi]\mapsto U^E(k_1)$ is plotted for $\mu=0.2\,,\,1.9\,,\,2.5$. The signature $\Sig(\widehat{T}^\mu_1)$ is $1,\,2,\,-1$ and the secondary invariant $\Sec(\widehat{T}^\mu_1)$ is $0,\,0,\,1$. In the two cases of odd signature there are Majorana edge modes.}
\label{fig-PPedge}
\end{center}
\end{figure}

%%%%%%%%%%%%%%%%%%%%%%%%%%%%%%%%%%%%%%%%%%%%%%%%%%%%%%%%%%%%%%%%%%%%

\end{document}